# Modeling and Evaluating Personas with Software Explainability Requirements

Henrique Ramos[1], Mateus Fonseca[2], Lesandro Ponciano[3]

[1] Software Engineering, PUC Minas, Belo Horizonte, Brazil

henriquealberone@outlook.com

[2] Software Engineering, PUC Minas, Belo Horizonte, Brazil

mateus.santos.fsc@gmail.com

[3] Dep. of Software Engineering and Information System, PUC Minas, Belo Horizonte, Brazil

lesandro.ponciano@gmail.com

**Abstract.** This work focuses on the context of software explainability, which is the production of software capable of explaining to users the dynamics that govern its internal functioning. User models that include information about their requirements and their perceptions of explainability are fundamental when building software with such capability. This study investigates the process of creating personas that include information about users' explainability perceptions and needs. The proposed approach is based on data collection with questionnaires, modeling of empathy maps, grouping the maps, generating personas from them and evaluation employing the Persona Perception Scale method. In an empirical study, personas are created from 61 users' response data to a questionnaire. The generated personas are evaluated by 60 users and 38 designers considering attributes of the Persona Perception Scale method. The results include a set of 5 distinct personas that users rate as representative of them at an average level of 3.7 out of 5, and designers rate as having quality 3.5 out of 5. The median rate is 4 out of 5 in the majority of criteria judged by users and designers. Both the personas and their creation and evaluation approach are contributions of this study to the design of software that satisfies the explainability requirement.

## 1 Introduction

The relationship between people and interactive systems has been an object of study in the area of Human-Computer Interaction [13, 18]. This area is interested in designing and evaluating interactive systems, considering the user, the interface, the interaction and the context of use [3, 13, 15]. The requirements of usability, accessibility, and communicability are precursor challenges in the design process in this area, which seeks to maximize the quality of the experience of its users from the early stages of system design to phenomena associated with use.

Over the last years, as interactive systems come to play a decisive role in the lives of individual people and in their collective behavior, new challenges have emerged in terms of requirements, usually described as new restrictions on system construction and functioning dynamics. Depending on how it is designed, interactive systems may inadvertently influence the opinions, choices and actions of their users, reflecting on social, political and economic dynamics. For example, when a system recommends a decision to the user over another or when it prioritizes some content over others without providing an explanation. In this context, it is increasingly required that this type of system be able to explain for a user its computation steps and how its outputs are generated [14]. It has been treated as a non-functional requirement, called "explainability requirement" [7,11].

Designing and implementing software so that it meets the explainability requirement is a major challenge. It is essential to understand to what extent people are concerned with explainability and to what extent they perceive the importance and feel the need for the system to be self-explainable. However, little is known about how users of interactive systems perceive this requirement. This work seeks to contribute to filling this gap by studying the process of modeling users including their perceptions, needs and concerns associated with the explainability requirement. In doing so, this study focuses on the technique of modeling users as personas, which are fictional characters created from real data to represent the target audience.

Our persona modeling approach integrates studies on the concept of explainability requirement [7], people's perception of the explainability requirement [11], creation of empathy maps and personas [6, 4], and evaluation of personas through the Persona Perception Scale [19]. It is a five-step process that can be summarized as follows: 1) questionnaires are applied to users to collect their perceptions and needs; 2) the responses obtained are used to create empathy maps including what the user says, feels, does and thinks about explainability; 3) similar empathy maps from different users are aggregated; 4) from the groups of empathy maps the personas are generated; 5) the personas are validated with the target public of users and designers. At the end of the fifth step, there is a set of personas that can be considered during the interface and interaction design so that the software may meet users' demands for explainability.

The proposed approach for persona creation and evaluation is investigated in this study in an empirical study with participation of 61 users in the first step (data collection), and 60 users and 38 designers in the fifth step (evaluation). The obtained results include a set of 5 distinct personas. Considering attributes of the Persona Perception Scale method [19], we found that personas are rated by the users as representative of them at an average level of 3.7 out of 5 and are rated by designers as having quality 3.5 out of 5. The median rate is 4 out of 5 in the majority of evaluation criteria. Both the personas and their creation and evaluation approach are contributions of this study for designers and researchers looking for strategies to guide the development of software with the explainability requirement.

The rest of this paper is organized as follows. We provide first a background of key concepts related to explainability and personas, and discuss relevant previous work (Section 2). Next, we discuss our approach to model and evaluate personas considering the explainability requirement (Section 3). After that, we detail the materials and methods of evaluation (Section 4). Then, we show and discuss the obtained results (Section 5). Finally, we discuss the conclusions of the study (Section 6).

## 2 Background and Related Work

This section presents the works related to the creation and use of user Empathy Map and Personas, as well as the advantages and disadvantages of its use. The section concludes with an analysis of the context of software explainability and recent advances in user modeling for that context.

As part of user modeling, this work uses *empathy maps*. Empathy Map is a user modeling technique that favors a better understanding of the user's context represented from 6 variables to be considered: what he says, does, sees, hears, feels and thinks. In addition to these, there are also areas of pain and need [5]. On the traditional Empathy Map, there are only the quadrants "thinks", "says", "feels" and "does", with the user represented in the middle. The first says about what the user thinks, but is not willing to vocalize. The "says" quadrant is what the user believes and that, if necessary, would speak without problems. Finally, the "feels" and "does" quadrants represent the user's feelings and attitudes, respectively. Empathy Map can be used to create personas [6].

A *persona* is a fictional character created to represent the target audience [9]. Its creation and use are relevant in User-Centered Design, a development approach in which the user must be understood during the entire process of conception, development and implementation of the product [8]. In this context, this model contains textual and graphic elements that incorporate the traits of target users. Personas help designers to have a more concrete view of who their users are [17]. In addition, they make the product developers sympathize with the people represented [2].

Even though the use of personas in collaborative design environments is well established, little research has been done to quantify the benefits of using this technique [10]. Previous results indicated that the groups of students who used personas produced products with superior usability characteristics. In addition, it is attested that the use of personas provides a significant advantage during the research and conceptualization stages of the design process. The fact that this study presents the advantages that the use of personas can bring to the users' experience makes it appropriate to be evaluated in the current study. Seeking to overcome limitations related to time and resources for data collection for the generation of personas, Mahamuni et al. (2018) evaluate the effectiveness of using the tacit knowledge of stakeholders in this process [12]. The use of tacit knowledge was effective in an organizational context, especially when time is a limitation.

Although personas are widely used in many domains, its evaluation is difficult, mainly due to the lack of validated measuring instruments. With this, the authors prepare a survey to assess the perception of individuals about a persona [19]. This artifact consists of 8 evaluation criteria, which can be modified to meet only those relevant to the research, each containing a maximum of 4 statements on a Likert scale:

1. **Credibility:** How realistic is the persona;
2. **Consistency:** The information in the description is consistent;
3. **Completeness:** Captures essential information about the described users;
4. **Clarity:** Information is presented clearly;
5. **Likability:** How nice the persona seems to be;
6. **Empathy:** How much the respondent empathizes with the persona;
7. **Similarity:** How much the persona looks like the respondent;
8. **Willingness:** Measures the respondent's willingness to learn more about the persona.

Applying the Perception Persona Scale [19], studies use clustering to validate automatically generated personas [1]. Among the criteria used in their validation survey are: similarity, empathy and credibility. Based on the results, it was noticed that two of the four generated personas achieved good results in the validation criteria, demonstrating that the participants have similar interests and think like the personas.

A useful way to understand the needs of users of a product or system is through the use of personas. However, Ferreira et al. (2015) consider that the creation of personas requires creativity and its validation, in terms of representativeness, is very difficult [4]. To assist in the creation of these models, the authors suggest the use of an empathy map. In the study, the designers' perception of the ease of use and usefulness of the empathy map for the creation of personas is assessed. To conduct the research, 20 user experience (UX) students participated, creating personas through textual content and then based on an empathy map. In line with this study, the work confirms that most designers found the Empathy Map technique easy to use and useful for creating personas.

In this work, personas were created in the context of software explainability. The challenge of developing software capable of explaining its outputs becomes greater the more sophisticated the computation performed by the software. One of the areas in which explainability has been widely addressed is the area of artificial intelligence, generally defined as Explainable Artificial Intelligence (XAI) [14,7]. An additional challenge in XAI is that the behavior of the software is not only dependent on its implementation, but also on the data used for training and learning the software. In the context of "deep learning" algorithms, there is a gap between the social meaning associated with users and the technical meaning associated with the implementation of the algorithms, which makes implementing the explanation even more challenging. Studies have shown that there are several gaps associated with the development of systems that adhere to the explainability requirement [14]. Studies on public participation systems have highlighted the importance of the explanation requirement being considered on system interaction with people [16].

In the context of user modeling contemplating information about the explainability requirement, a technique that has already been considered is profile [11]. Louzada et al. (2020) seek to identify similarities and differences between users of interactive systems in terms of the importance of the requirement of software explainability, using the profile technique based on clustering. The study found 6 profiles, each with their level of interest in explainability of an interactive system. The study discusses the importance of creating personas and motivates further studies in this direction. Thus, this paper helps to advance this literature on this aspect of user modeling by including information on demand for explainability.

## 3 An approach for modeling and evaluating personas considering the explainability requirement

In this section, we present our approach for modeling and evaluating personas that include information about how users perceive the explainability requirement and their needs regarding such requirements. In general, this approach integrates studies on the concept of explainability requirement [7], people's perception of the explainability requirement [11], creation of empathy maps and persona [6, 4], and evaluation of personas through the Persona Perception Scale [19]. The proposed approach for creating and evaluating personas has 5 steps as summarized below:

1. questionnaires are applied to users in order to collect their perceptions and needs about explainability;
2. the obtained answers are used to automatically create the four dimensions of an empathy map for each user;
3. users with similar empathy map are aggregated on only one group of empathy maps;
4. from the resulting groups of empathy maps, the personas are generated;
5. Personas are validated with the public of users and designers.

The questionnaire used in the **first step** to analyze perceptions and explanatory needs is based on the questionnaire proposed by Louzada et. al (2020). The questionnaire investigates needs and requirements on explainability (Table 1). Depending on each context of use, other demographic information may be included, such as age, gender and schooling. The public to which the questionnaire is applied has a wide effect on the personas that will be obtained. For example, if the public are mostly male, at the end of the process, the results tend to be more personas with this characteristic. In the **second step**, the questions in this questionnaire are mapped on the dimensions of the empathy map [4], as shown in Table 1. The answers to each of these questions are on the 5-point Likert scale, being coded in answers from 1 to 5. For the quadrants "feels", "thinks" and "says", the subtraction of the answer pair, for being contradictory issues. For the "does" quadrant, the average of the answer pair is calculated. In the four dimensions, if the result is greater than or equal to 2.5, it is classified as "positive", otherwise, as "negative". Thus, at the end of the second step, for each user, there are four dimensions and each of them has a value defined as positive or negative.

**Table 1.** Questions used to derive the quadrants from the empathy maps. Items in the "think", "feel" and "say" quadrants are answered in a five-point Likert scale. In the quadrant "does" answers are options that refer to acting positively or negatively in relation to the software explanations.

| Quadrant | First question | Second question |
|---|---|---|
| DOES | Suppose you are using software where you enter the address of the location you are at and the address of the location you want to go to and the software tells you which street path you must follow to get to the desired location. Select the option that most closely matches your behavior in this situation. | Suppose you are using software that is a social network where you can follow people and be followed. Suppose also that the software recommends someone to you to follow. Select the option that most closely matches your behavior in this situation. |
| THINKS | If a user is interested in knowing how software generates recommendations that it makes, so the software must provide such an explanation to that user. | I follow a software generated recommendation if it is useful to me, regardless of whether or not it has an explanation associated with it. |
| FEELS | I feel more confident in following a recommendation made by a software when it explains to me why it considers the recommendation suitable for me | I usually feel confused by recommendations that the software I use makes me when they are not explained. |
| SAYS | Software should be required by law to provide explanations of how they generate the recommendations they present to users. | I have no interest in knowing how the software I use generates recommendations for me |

In the **third step**, the empathy maps are grouped. For example, if 4 users have "positive" value in all 4 dimensions of the empathy map, then these users are equal and can be represented by only 1. As a result of this process, there is at least 1 empathy map, if all users are equal, and at most 16 empathy maps, which is the case that there are the two combinations of values (positive and negative) in each of the four quadrants of the empathy maps, so $2^4 = 16$ possibilities.

In the **fourth step** the grouped empathy maps are transformed into personas. For this, there are two fundamental activities. The first activity is to identify the demographic characteristics of the participants. This is done for each group, in which the modal value of age, gender, education and other user characteristics of users in each group is obtained. The second activity is to interpret the quadrants of the group's empathy map to describe it as an element of the persona. This interpretation is done following Table 2.

**Table 2.** Positive and negative interpretations per quadrant of the Empathy Map.

| Quadrant | Positive rating | Negative rating |
|---|---|---|
| DOES | Tends to follow the recommendation provided by the software. | Tends not to follow the recommendation, makes his decisions alone. |
| THINKS | Tends to believe that systems should explain its recommendations. | Tends not to care about software explanations of its recommendations. |
| FEELS | Feels more comfortable following a well-explained recommendation. | A well-explained recommendation does not change his decision to follow it. |
| SAYS | Says that explanations must be provided to users who are interested. | It says that explanations should not be obligatorily provided. |

In the **fifth step**, two questionnaires with questions from Persona Perception Scale [19] with answers in Likert scale, between 1 (I totally disagree) and 5 (I totally agree), are applied. The questionnaire available in Table 3 is applied to users to quantify their perception of representation, taking into account the criteria: similarity, empathy and sympathy. The questionnaire available in Table 4 is applied to designers to measure their perception of the quality of the artifact, taking into account its clarity, completeness and credibility.

**Table 3.** Questionnaire about user perception of personas representativeness from Persona Perception Scale. The items are answered in a five-point Likert scale.

| Construct | Item |
|---|---|
| Similarity | This persona feels similar to myself. |
| Similarity | The persona and I think alike. |
| Similarity | The persona and I share similar interests. |
| Similarity | I believe I would agree with this persona on most matters. |
| Empathy | I feel like I understand this persona. |
| Empathy | I feel strong ties to this persona. |
| Empathy | I can imagine a day in the life of this persona. |
| Likability | I find this persona likable. |
| Likability | I could be friends with this persona. |
| Likability | This persona is interesting. |
| Likability | This persona feels like someone I could spend time with. |

**Table 4.** Questionnaire about designer perception of personas quality from Persona Perception Scale. The items are answered in a five-point Likert scale

| Construct | Item |
|---|---|
| Credibility | Those personas seem like real people. |
| Credibility | I have met people like those personas. |
| Credibility | The picture of those personas looks authentic. |
| Credibility | Those personas seem to have a personality. |
| Completeness | Those personas profiles are detailed enough to make. decisions about the customers they describe. |
| Completeness | Those personas profiles seem complete. |
| Completeness | Those personas profiles provide enough information to understand the people they describe. |
| Completeness | Those personas profiles are not missing vital information. |
| Clarity | The information about the personas is well presented. |
| Clarity | The text in the persona's profile is clear enough to read. |
| Clarity | The information in the persona's profile is easy to understand. |
| Clarity | Those personas are memorable. |

Both questionnaires applied have a set of personas. In the questionnaire on the perception of users shown in Table 3, the user must select which persona most represents her/him and then answer the questions, taking into account only the selected persona. In the questionnaire on the perception of designers, shown in Table 4, designers must evaluate each group of personas as a whole.

## 4  Materials and Methods of Evaluation

This work seeks to analyze the quality of personas in relation to the representation of users and their construction in the context of software explainability, taking into account their creation through the approach proposed in the previous section. The research carried out is a case study in which the proposed approach was followed from its first step to fifth step. It is a case study because the study is carried out with a specific audience, although the proposed method can be applied to other audiences. This section describes the materials and methods employed in such a case study.

All questionnaires applied in this study, whose questions are described in section 3 are prepared in Google Forms. The questionnaire about perception and needs of explainability (first step) was applied to 61 people in 2020; they are mostly members of laboratories and technology development companies. The personas evaluation questionnaires (fifth step) were applied between April 13, 2021 and April 24, 2021. There were 38 participants who acted by answering the questionnaire as designers, being people who work in design interaction teams in software development companies in the city of Belo Horizonte, Minas Gerais state, in Brazil. There were 60 participants who responded to the questionnaire as users, being people with the same characteristics as the target audience consulted in the first step, in which the needs and perceptions of explainability were collected.

In the user perception questionnaire, the participant sees a set of personas and the questions are answered taking into account the persona with which the participant most identifies. In the questionnaire on the perception of designers, the set of personas is evaluated as a whole. Analysis is always done separately for users and designers.

The results reported in this study are the participant's level of agreement for the Persona Perception Scale items and the average agreement. Participant's level of agreement is quantified from 1 to 5. The higher the value, the more participants agree with the evaluated item. The average agreement is calculated by summing the level of agreement from all participants and dividing by the number of participants. The higher the average value, the more participants agree with the evaluated item.

The average agreement metric is calculated in two different scenarios: average agreement by construct and overall average agreement. In the average agreement per construct, the average agreement is grouped by the assessment construct defined in the Persona Perception Scale, thus, for each respondent, there is an assessment value per construct and the average obtained from the set of respondents is reported. In the overall average agreement, there is the general value for each participant and the reported average is the overall average value in this set of participants.

In all results reported in this study, the error bars are shown for a statistical error at a confidence level of 95%, being the calculations performed by using the R-statistics language.

## 5 Results

In this section, we detail the results of our case study in generating personas that include information about users' perceptions and needs of software explanations. In doing so, we first present the set of personas generated by using the approach described in section 3. After that, we analyze the perceptions from users and from designers about representativeness and the quality of the personas. Finally, we discuss the results of the distribution of responses of participants per item of the Persona Perception Scale questionnaire. Figure 1 shows the five personas generated in the case study.

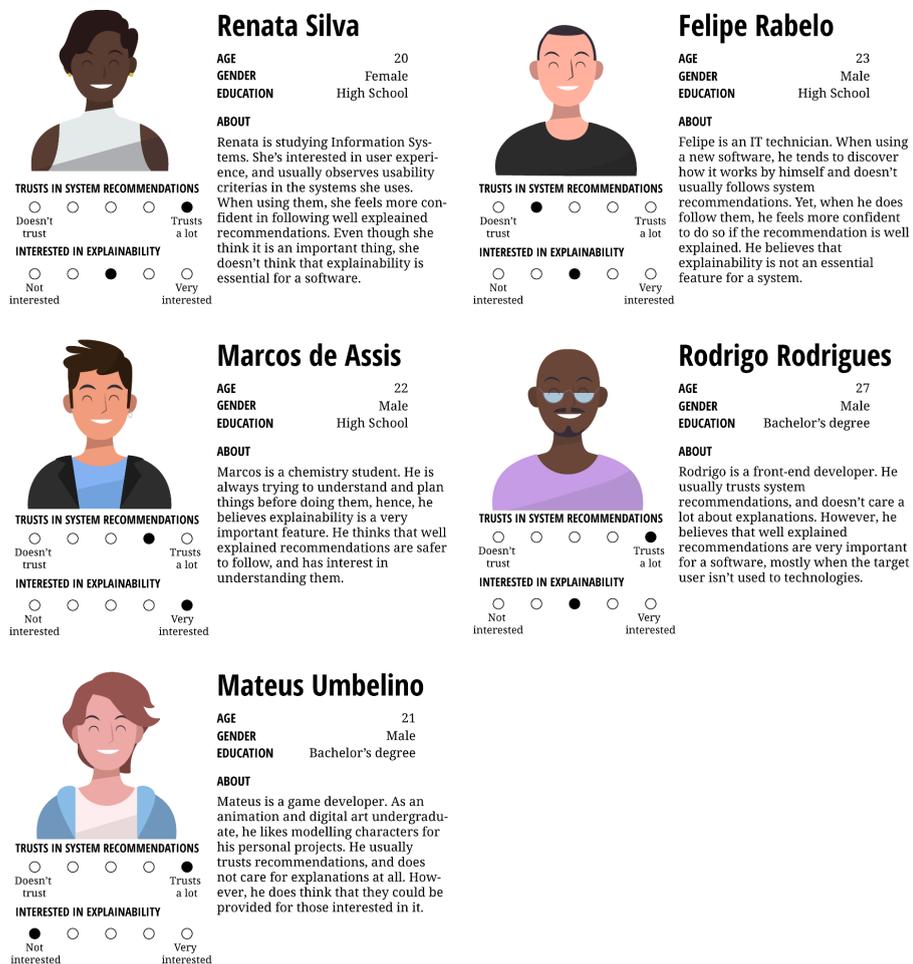

**Fig. 1.** Personas generated from the responses of 61 participants considering their needs and perceptions of software explainability, aggregated empathy maps, and demographic data.

As discussed earlier, personas are produced from the aggregation of the empathy maps created with the responses from the perception and need for explainability questionnaire. As shown in Figure 2, considering the public of 61 respondents, the personas were originated as follows: Marcos Assis (34% of respondents), Renata Silva (23% of respondents), Mateus Umbelino (18% of respondents), Rodrigo Rodrigues (17% of respondents), and Felipe Rabelo (8% of respondents). Thus, in addition to modeling different types of users, there are personas that are more common in the target audience and others that are less common.

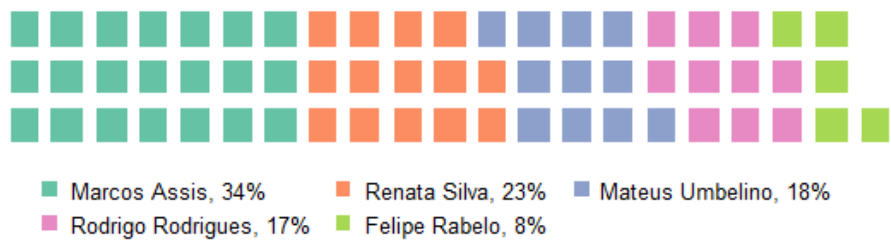

**Fig. 2.** Number of respondents whose empathy maps were aggregated and originated each of the personas.

Figure 3 shows results of the average agreement of users, per construct considered in the Persona Perception Scale. On the X-axis are the construct (Similarity, Empathy and Likability) and also the overall case, which includes all constructs together. The results show an overall rating of 3.7 on users' average agreement. There was more agreement on the Likability construct, indicating that the respondents like the personas.

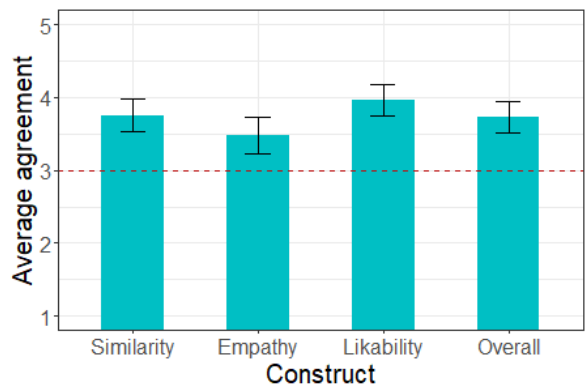

**Fig. 3.** Users' average level of agreement to the criteria of representativeness and quality. The dashed line is the level of agreement equivalent to "neither agree nor disagree" Each error bar represents the 95% confidence interval.

Figure 4 shows results of the average agreement of the designers, per construct considered in the Persona Perception Scale. On the X-axis are the construct (Credibility, Completeness and Clarity) and also the overall case, which includes all constructs together. The results show an overall rating of 3.5 on designers' average agreement. There was more agreement on the Clearness construct and less agreement in the Completeness, indicating that the designers perceive the personas as succinct and direct, but not complete. This is an expected result, as the personas seek to contemplate the explainability requirement but not cover information relevant to other domains that are not relevant in this domain.

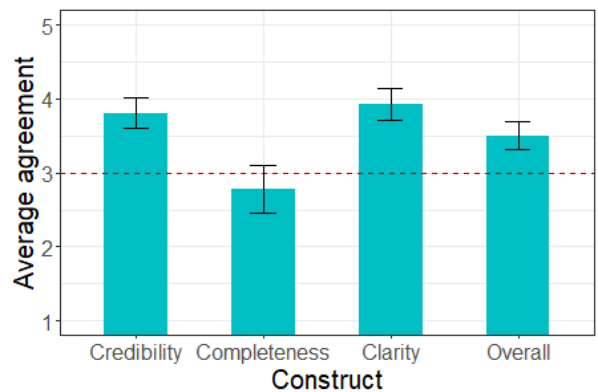

**Fig. 4.** Designers' average level of agreement to the criteria of representativeness and quality. The dashed line is the level of agreement equivalent to "neither agree nor disagree" Each error bar represents the 95% confidence interval.

Figures 5 and 6 show the agreement distribution of both designers and users regarding personas. In them, the Y axes are the questions of the questionnaire on the perception of representativeness, applied to users, and the questionnaire on the perception of quality, applied to designers, respectively, while the X axis represents the level of agreement.

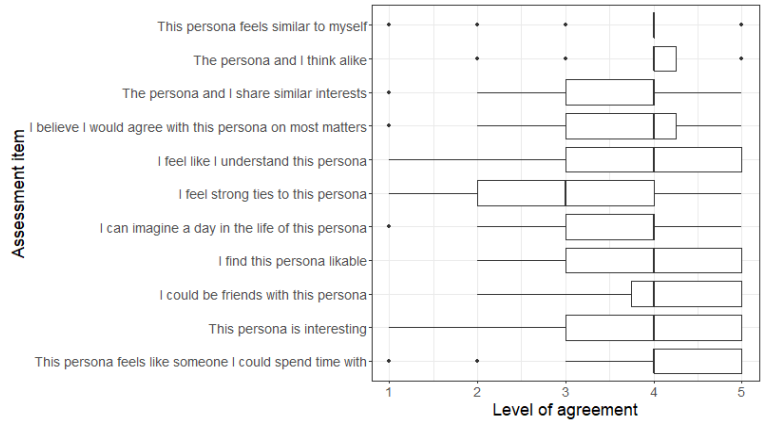

**Fig. 5.** Users' perception. Distribution of responses from users about the personas, considering items of the Persona Perception Scale method.

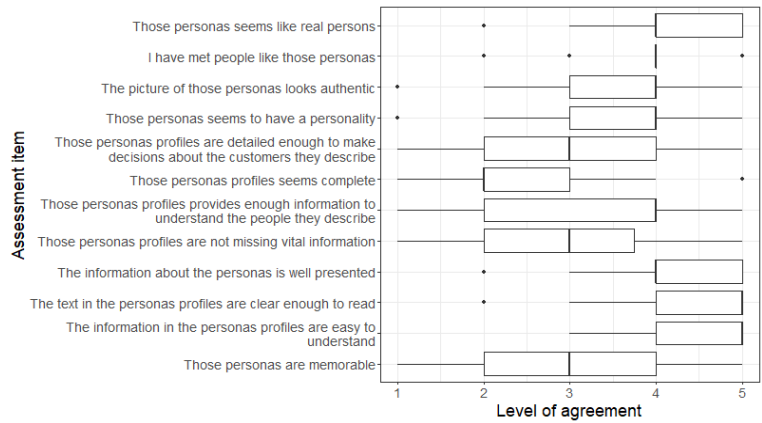

**Fig. 6.** Designers' perception. Distribution of responses from designers about the personas, considering items of the Persona Perception Scale method.

This result shows that most interquartile ranges are positioned between 3 and 5, this means that the assessment of at least 50% of the questionnaire items is greater than or equal to 3, that is, above the average of 2.5. In addition, the medians of both graphs also present evaluations above the average in most, except in Figure 6, where the evaluation items that correspond to the completeness criterion mix medians between 2 and 4. Furthermore, the questionnaire outliers are, predominantly, of evaluations below 3, showing that these evaluations are minority.

Finally, the statements "This persona feels similar to myself" and "I have met people like this persona", belonging to the similarity and credibility criteria, respectively, had the values of median, 25th percentile and 75th percentile in 4, indicating that the majority of the participants found similarities of the personas with themselves or with people close to them. In contrast, the statements "I feel strong ties to this persona" and "This persona is memorable", which belong to the criteria of empathy and clarity, respectively, have responses in a practically normal distribution, indicating total convergence between the central tendency measures.

## 6   Conclusion

In this work, we focused on the context of software explainability, which is the production of software capable of explaining to users the dynamics that govern its internal functioning. An approach of creating user models that include information about their requirements and their perceptions of explainability are fundamental when building software with such capability. So, our study investigated an approach of creating personas that include information about users' explainability perceptions. The proposed approach has five steps as follows: 1) questionnaires to collect users' perceptions and needs on the context of explainability; 2) the responses obtained are used to create empathy maps; 3) such maps are grouped by similarity; 4) from the groups the personas are generated; 5) the personas are validated with the public of users and designers.

In a case study, we employ the approach with the participation of 61 users. The obtained results include a set of 5 distinct personas representing mostly members of laboratories and technology development companies. A public of 60 users and 38 designers participate in the evaluation. The personas were rated by the users as representative of them at an average level of 3.7 out of 5 and are rated by designers as having quality 3.5 out of 5. The median rate is 4 out of 5 in most evaluation criteria, for both users and designers. We believe that both the personas and their creation and evaluation approach are relevant contributions for designers and researchers looking for strategies to guide the development of software that satisfies the explainability requirement.

Several future works can be conducted based on what is presented in the study. In particular, we plan to employ personas in the software development process, that is, to investigate how designed people will build different interfaces and interactions from them. We also intend to seek the creation of more inclusive personas, by carrying out data collection (first step of the approach) with a wider and more diverse audience in terms of gender, race, geographic region and also the so-called Extreme Characters. Thus, this study can support and motivate further work in the context of creating and using user models to build software with the requirement of explainability.

## References

1. Branco, K.d.S.C, Oliveira, R.A., Silva, F.L.Q.d, de H. Rabelo, J., Marques, A.B.S.: Does this persona represent me? investigating an approach for automatic generation of personas based on questionnaires and clustering. In: Proceedings of the 19th Brazilian Symposium on Human Factors in Computing Systems. IHC'20, ACM, New York, NY, USA (2020). https://doi.org/10.1145/3424953.3426648

2. Cooper, A., Saffo, P.: The Inmates Are Running the Asylum. Macmillan Publishing Co., Inc., USA (1999)
3. De Souza, C.S.: The semiotic engineering of human-computer interaction. MIT press (2005)
4. Ferreira, B., Silva, W., Jr., E.A.O., Conte, T.: Designing Personas with Empathy Map. In: Proceeding of the 27th International Conference on Software Engineering and Knowledge Engineering. pp. 501-505. KSI Research Inc. and Knowledge Systems Institute Graduate School, Pittsburgh, USA (2015). https://doi.org/10.18293/SEKE2015-152
5. Gasca, J., Zaragozá, R.: Designpedia. 80 herramientas para construir tus ideas. LEO, LID Editorial Empresarial, S.L. (2014)
6. Junior, P.T.A., Filgueiras, L.V.L.: User modeling with personas. In: Proceedings of the 2005 Latin American Conference on Human-Computer Interaction. p. 277–282. CLIHC '05, ACM, New York, NY, USA (2005). https://doi.org/10.1145/1111360.1111388
7. Köhl, M.A., Baum, K., Langer, M., Oster, D., Speith, T., Bohlender, D.: Explainability as a non-functional requirement. In: 2019 IEEE 27th International Requirements Engineering Conference (RE). pp. 363–368 (2019). https://doi.org/10.1109/RE.2019.00046
8. LaRoche, C.S., Traynor, B.: User-centered design (ucd) and technical communication: The inevitable marriage. In: 2010 IEEE International Professional Communication Conference. pp. 113–116 (2010) . https://doi.org/10.1109/IPCC.2010.5529821
9. Lidwell, W., Holden, K., Butler, J.: Universal principles of design, vol. 1. Rockport, Gloucester, MA (2010)
10. Long, F.: Real or imaginary; the effectiveness of using personas in product design. In: Proceedings of the Irish Ergonomics Society Annual Conference. p. 1–10. Dublin (05 2009)
11. Louzada, H., Chaves, G., Ponciano, L.: Exploring user profiles based on their explainability requirements in interactive systems. In: Proceedings of the 19th Brazilian Symposium on Human Factors in Computing Systems. IHC'20, ACM, New York, NY, USA (2020). https://doi.org/10.1145/3424953.3426545
12. Mahamuni, R., Khambete, P., Punekar, R.M., Lobo, S., Sharma, S., Hirom, U.: Concise personas based on tacit knowledge - how representative are they? In: Proceedings of the 9th Indian Conference on Human Computer Interaction. p. 53–62. IndiaHCI'18, ACM, New York, NY, USA (2018). https://doi.org/10.1145/3297121.3297126
13. Nielsen, J.: Usability engineering. Elsevier (1994)
14. Nunes, I., Jannach, D.: A systematic review and taxonomy of explanations in decision support and recommender systems. User Modeling and User-Adapted Interaction 27(3), 393–444 (2017). https://doi.org/10.1007/s11257-017-9195-0
15. Ponciano, L., Brasileiro, F., Andrade, N., Sampaio, L.: Considering human aspects on strategies for designing and managing distributed human computation. Journalof Internet Services and Applications 5(1), 1–15 (2014). https://doi.org/10.1186/s13174-014-0010-4
16. Ponciano, L., Pereira, T.E.: Characterising volunteers' task execution patterns across projects on multi-project citizen science platforms. In: Proceedings of the 18th Brazilian Symposium on Human Factors in Computing Systems. IHC'19, ACM, New York, NY, USA (2019). https://doi.org/10.1145/3357155.3358441
17. Pruitt, J., Adlin, T.: The Persona Lifecycle: Keeping People in Mind Throughout Product Design. Morgan Kaufmann Publishers Inc., San Francisco, CA, USA (2005)
18. Rogers, Y., Sharp, H., Preece, J.: Interaction design: beyond human-computer interaction. John Wiley & Sons (2011)
19. Salminen, J., Santos, J.M., Kwak, H., An, J., gyo Jung, S., Jansen, B.J.: Persona perception scale: Development and exploratory validation of an instrument for evaluating individuals' perceptions of personas. International Journal of Human-Computer Studies 141, 102437 (2020). https://doi.org/10.1016/j.ijhcs.2020.102437